# CONTEXT-BASED BARRIER NOTIFICATION SERVICE TOWARD OUTDOOR SUPPORT FOR THE ELDERLY

Keisuke Umezu[1], Takahiro Kawamura[2] and Akihiko Ohsuga[2]

[1]NEC Corporation
[2]Graduate School of Information Systems, University of Electro-Communications, Japan

## ABSTRACT

*Aging society has been becoming a global problem not only in advanced countries. Under such circumstances, it is said that participation of elderly people in social activities is highly desirable from various perspectives including decrease of social welfare costs. Thus, we propose a mobile service that notifies barrier information nearby users outside to lowers the anxiety of elderly people and promote their social activities. There are barrier free maps in some areas, but those are static and updated annually at the earliest. However, there exist temporary barriers like road repairing and parked bicycles, and also every barrier is not for every elder person. That is, the elder people are under several conditions and wills to go out, so that a barrier for an elder person is not necessarily the one for the other. Therefore, we first collect the barrier information in the user participatory manner and select the ones the user need to know, then timely provide them via a mobile phone equipped with GPS. This paper shows the public experiment that we conducted in Tokyo, and confirms the usability and the accuracy of the information filtering.*

## 1. INTRODUCTION

Recently, the rapid aging of society has emerged as a global problem. In Japan, elderly people (over 65 years of age) accounted for more than 23.1% of the total population in 2010 [1]. In this inevitable trend in the world, participation of the elderly people in social activities is getting much more important from various perspectives including employment coordination and social welfare, medical costs. However, the elderly people have several anxieties in contemporary society. Actually, the people with any of disabilities have to make a great deal of effort in order to move around outside their homes. This leads to rarely leaving the house. In some areas there are barrier free maps provides, but those are big, static maps and updated annually at the earliest. In reality, there exists temporary barriers like road repairing and parked bicycles, and also every barrier is not for every elder person. That is, the elder people have several different conditions and wills to go out, so that a barrier for an elder person is not necessarily the one for the other.

Therefore, we propose a mobile service for the elderly that employs cellular phones equipped with GPS. It collects the barrier information in the user participatory manner. Also, it automatically selects the barrier information in an urban environment according to the user's context, and then notifies it to the user via the cellular phone. Then, we evaluate the usability regarding the burden of operating the service and the accuracy of the barrier notification corresponding to the user context.

The remainder of this paper is organized as follows. Firstly, section 2 refers to the related works, and then section 3 defines requirements of the service for the elderly on the basis of the results of our survey. Then, the service overview and the barrier notification mechanism are described in section 4. Sections 5 and 6 present the experiments and results with regard to the usability and the notification accuracy, respectively. Finally, we present our conclusions in section 7.

## 2. RELATED WORKS

Sakamura et al. conducted an experiment using Ubiquitous Communicator [2]. It is a handheld device that reads RFID tags embedded at several places in a town, and shows useful information on the place nearby. The RFID has advantage in the indoor situation where GPS can not take the position of the user. However, for the outdoor use the cost of embedding RFID tags throughout an area will pose a problem for the practical use.

Yairi et al. created barrier-free maps for a few towns and launched a commercial service [3]. Users use their PCs to browse Web-based online maps and they can input barriers. Also, it can create a route to a goal considering the barriers. A similar approach can be found at [4], which is accessible from the smartphone. However, their interfaces are limited to showing a map including every spot. On the other hand, our service has improved its usability for the elder users, and especially featured by the content selection mechanism.

Kurihara et al. are using a high-resolution GPS whose error is less than a few centimetres and trying to create a barrier-free map automatically [5]. In this project, the user's route is extracted by the GPS attached to a wheelchair. Then, steps and flatness are automatically detected by the trace. We would like to consider the use of that GPS, however in this paper we selected a more commodity device, the GPS-equipped cellular phone, because it is already in widespread use among the elder people.

With regard to a social bookmarking function in the real world (content submission), there are several researches such as Space Tag [6], xExplorer [7], Voting With Your Feet [8], Place-Its [9]. In those researches, GPS is used not only for the information notification, but also for users' communication and reminder systems. In the future, we will consider incorporation of those features in order to promote the use of the content submission mechanism among the elderly.

## 3. REQUIREMENTS OF MOBILE SERVICE FOR THE ELDERLY

### 3.1. Survey for problem in Elderly People

In this research area, there are several discussions based on someone's imagination, which are not necessarily tackling the practical problem. Therefore, we conducted in advance a survey whose subjects were the authors' neighbours, relatives and members of the Japan National Council of Social Welfare. We then collected their problems and complaints concerning daily life. The followings are the extracted problems we would like to focus in this paper.

1. Decrease of the willingness to leave the home
The main factors decreasing the willingness of the elderly to leave the house are barriers such as pedestrian ways with steps and stairs, illegally parked cars and bicycles, and lack of sidewalks.
2. Desire to be independent
The elderly are grateful for the support from their families and volunteers, but most of them are reluctant to accept it.
3. Difficulty in using new machines

The elderly tend not to be good at using appliances and computer systems, or are unfamiliar with them.

### 3.2. Service Requirements for Elderly People

To solve these problems, we extracted the following three requirements.

1.  Barrier / useful information notification

Although the ultimate solution is to remove the barriers, we propose to provide the users who go out with notification of the existence of barriers, and then indicate alternative routes to avoid the barriers for reduction of the anxiety. Of course, the notified content is not restricted to barriers, and useful information such as the location of public rest rooms, slopes and places of interest would also motivate the elderly to go out.

2.  Easy-to-use interface

The system is required to be easily operated by the elderly. Therefore, when operation is necessary, interfaces such as buttons should be of an appropriate size and well designed. Icons would be useful, too.

3.  Autonomous behaviour

Autonomous behaviour is also important for the elderly who tend to find information systems difficult to use. The current navigation systems for pedestrians such as NAVITIME [10] require active operations by the users, i.e., goal input, route choice, and symbol selection, etc. In a system for the elderly, however, the service should actively provide information to passive users.

## 4. BARRIER NOTIFICATION SERVICE

To satisfy the above requirements and promote participation in social activities, we have developed a mobile service using a GPS cellular phone which supports for the elderly to go out with less anxiety. We selected the cellular phone equipped with GPS as the service client, but it is not a smartphone, because their accessibility based on tapping operation had negative feedback from the elders. But, GPS function is prerequisite because 3G cellular phones have been required to be equipped with GPS from 2007 in Japan. Note that our target users in this research are people who are physically able to go out, but tend to stay at home [1]. Disabled people who need physical support are out of the scope of this paper.

### 4.1. Service overview

Fig. 1 shows the service flow. First of all, the user starts the service on the phone, and keeps it running before going out. When the user comes close to a spot where the barrier or useful information is registered in a SNBI (Searching NearBy Information) server, our dog character "CoCo" (means "Here" in Japanese) barks "bow wow" to alert the user, and displays its content composed of a picture and a brief description (Fig. 2). We set a large character size and high sound volume considering the impaired vision and hearing experienced by many of the elderly. If the user pushes the orange button when notified, a map of the area indicating the location of the content is displayed. Of course we can show all the nearby content at once on a map, but the detailed map on the small screen and operations on the map would not good for the elderly.

Also, the system provides a content submission mechanism. It is like a human computation in the real world. It is practically impossible for the public administration to update all the contents in the world everyday. So that a social approach among individual users is a key for the content collection in the future. For the content submission, the user pushes a circular button for useful information or a cross button for a barrier at the location (Fig. 3). Then, the built-in camera is activated (currently, 90% of cellular phones sold in Japan are equipped with a digital camera), and thus the user can input title, comment, tags, and time such as 9am-5pm after taking a photo. After completing the inputs, the user pushes the send button to register the information and the current location to the SNBI server. The content becomes available for all subsequent users who approach that location.

The service is realized by combination of a Java application on cellular phones and Web services. The SNBI server is implemented as a Java servlet. If the client queries the SNBI server to get the nearest content, the GPS module is automatically invoked and the retrieved location information is sent to the SNBI server. Then, the SNBI server searches the DB (content table) based on the current location, and returns the nearby content, if found. If not, the map including the current location is sent to the client. This is repeated every 2 to 3 minutes. The locations are stored in another DB (trace log table) in the SNBI server, and are used to trace and predict the route.

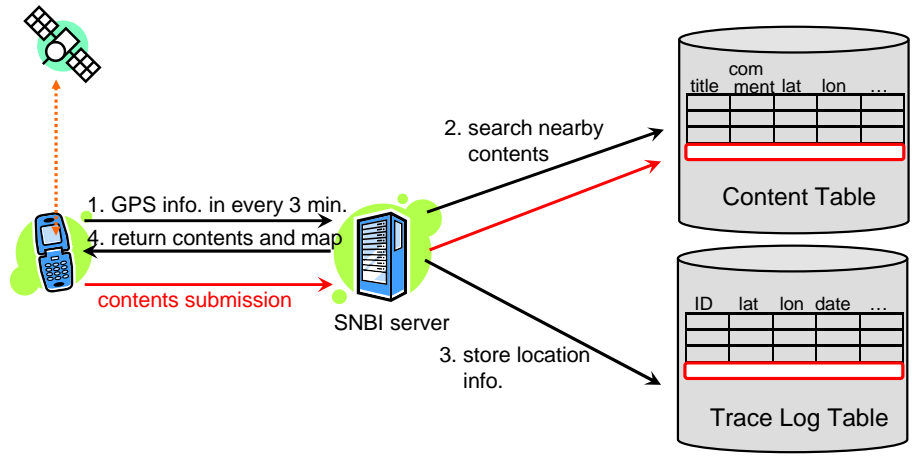

Figure 1. Service flow

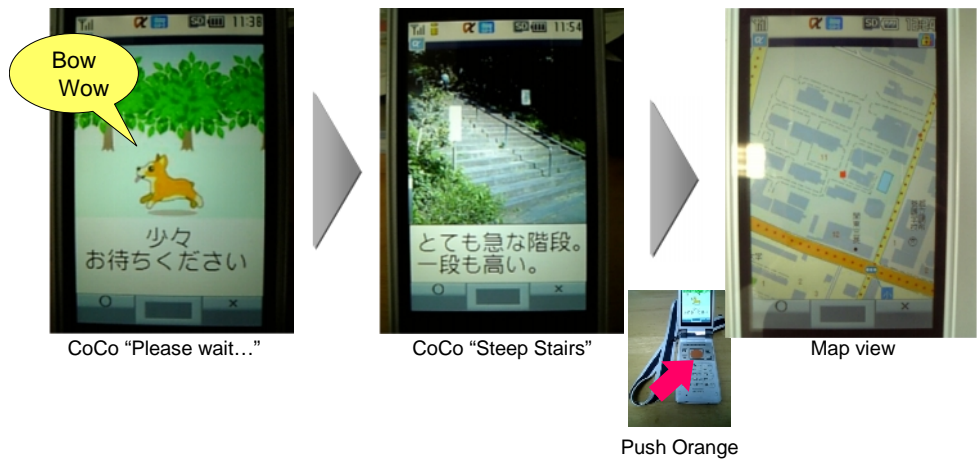

Figure 2. Screen transition for notification

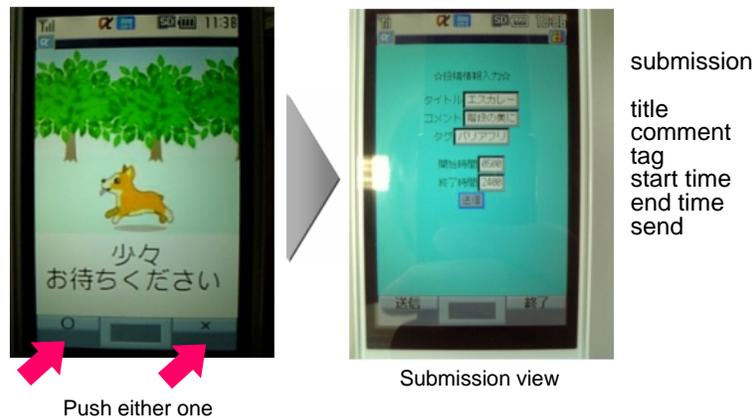

Figure 3. Screen transition for submission

## 4.2. Content notification mechanism

Content selection is a key issue of this service. The elder users are under several conditions. That is, one might have difficulty on the knee, and another may have weak hearing. Also, their will to go out is ranging from rehabilitation to change. So if we emit every content (barrier and useful information) to the user, it can easily overwhelm the user and counteract the will to go. Or, the service will be dumped. Therefore, the notification of unnecessary content for the user must be reduced. At the same time, there are several ways around the barriers. For example, in the case of a pedestrian bridge, there are two ways: detour (finding a crossing), or across the bridge. Thus the notification mechanism should propose a proper approach according to the user's profile. Furthermore if the way around is "across the bridge", the "bridge" might not be a barrier for the user and it should be recorded.

In this paper, we used a Bayesian network [11] to select the content. The Bayesian network estimates probabilistic parameters from a given statistical data set and a causal association model, and can be used to predict cases with uncertainty [12]. For the content selection, it's practically impossible to collect every case for every profile and make if-then rules. Therefore, we used the Bayesian network which estimates probability distribution even for missing data sets and reasons unseen cases. Fig. 4 shows our Bayesian network model and parameters for the content selection. It has environmental parameters (weather, temperature, etc.) and the user profile (purpose of going out, familiarity of the place, ability to walk a long distance, willingness to walk), and then predicts the importance of each content for the user after the direction filtering. The direction filtering is done by selecting contents located within 100 degrees ahead on the user's direction which is simply a straight line from previous position to current position (see Fig. 5).

According to this content selection based on the user profile, we would be able to do the followings: if the user does not know the place well, it would be important to check the nearest police box. But, if the user goes there frequently, such notification is unnecessary. Furthermore, an escalator is of course important for a disabled user, but stairs should be recommended to the user who tends to walk for exercise.

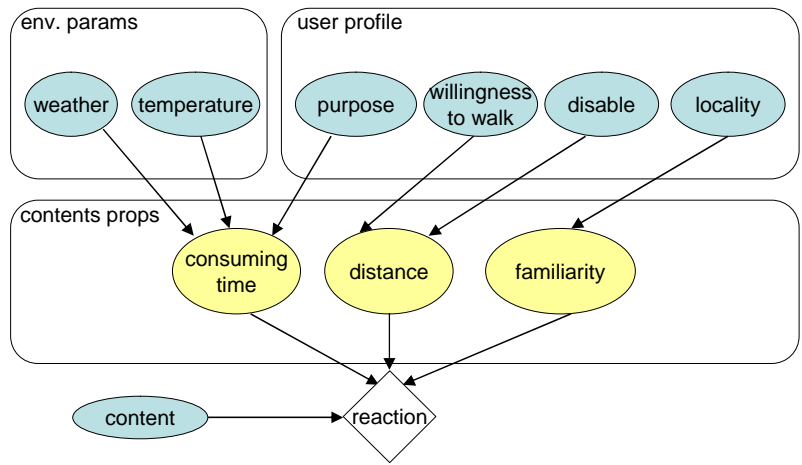

Figure 4. Bayesian network for content selection

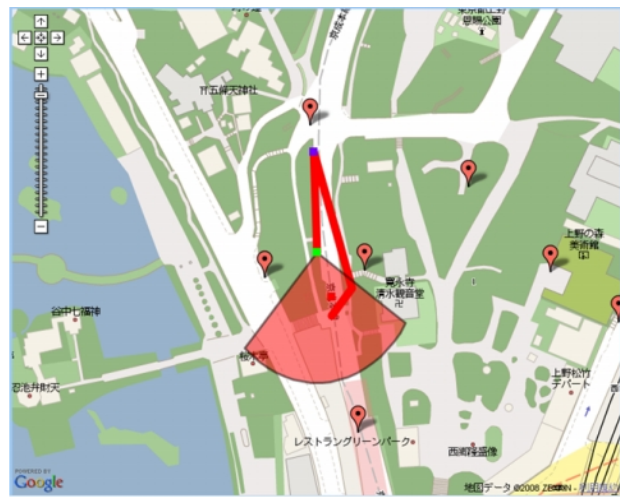

Figure 5. Direction filtering

## 5. EXPERIMENT ON USABILITY

### 5.1. Experiment Setting

Firstly, to check the usability of our system which corresponds to the above requirement 2, 3 we conducted an experiment at Ueno Park, Tokyo, for 10 days in August. Test users are randomly selected from the elderly people walking in the park (Fig. 6). Since this area includes museums, ponds and many other places of interest, it is popular with the elderly. Before the experiment, we collected 35 contents of the barriers and useful information, and then registered them in advance, such as the places where street people and crows gather, stairs and steps, and benches in the shade.

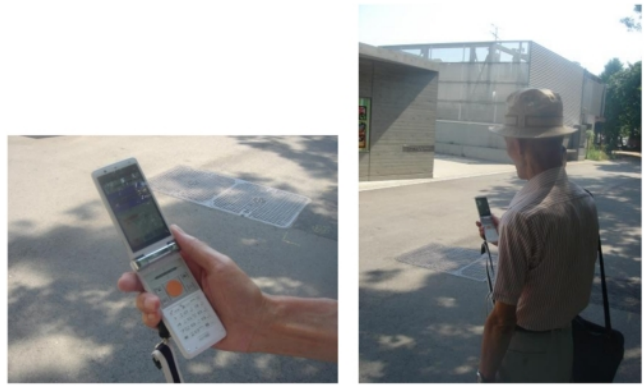

Figure 6. Snapshot of experiment

At the experiment, we first explained the purpose of this experiment for each user and instructed the use of the client, that is, how to get notification and how to submit content. Then, it was conducted from the user's current position to the user's goal location, and then we orally collected the responses to a questionnaire for 10 min. As the result, we obtained the walk routes (Fig. 7) and the responses to the questionnaire from 10 pairs of elderly people (Table 1).

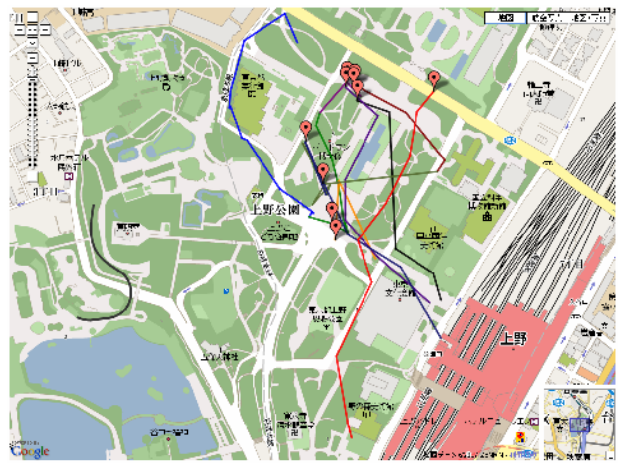

Figure 7. Walk routes of 10 users at Ueno Park

Table 1. Profiles of test users

| ID | Sex, Num | Age | Freq. of visit | Disability | Stairs and Steps | Have a Cell Phone? |
|----|----------|-----|----------------|------------|------------------|--------------------|
| 1  | F, 2p    | 70s | Several times  | Legs       | Never            | No                 |
| 2  | F, 2p    | 70s | 2 to 3         | Legs       | Never            | No                 |
| 3  | M        | 60s | 2 to 3         | Legs       | Never            | Yes                |
| 4  | M & F    | 80s | Several times  | None       | Prefer not to    | Yes                |
| 5  | F, 2p    | 80s | Several times  | None       | Prefer not to    | No                 |
| 6  | M & F    | 70s | 2 to 3         | None       | Prefer not to    | No                 |
| 7  | M & F    | 60s | First time     | None       | OK               | No                 |
| 8  | M, 2p    | 70s | Several times  | None       | OK               | No                 |
| 9  | M & F    | 70s | Several times  | None       | OK               | Yes                |
| 10 | M        | 80s | Several times  | None       | OK               | No                 |

## 5.2. Experiment Results

### 5.2.1. Contents Notification

In terms of notification timing along with the user's position, the notified content was:

- in front of the user's direction (8 cases, 24%),
- at the same location as the user (10 cases, 30%),
- behind the user's direction (8 cases, 24%),
- misaligned such as on another street (7 cases, 21%).

Also, in their responses to the questionnaire, half of the test users complained that the notifications were not provided at the right time (Fig. 8). We will improve this point by estimating the user's walking speed and direction. Notification of content along with the user's position is discussed in the next section.

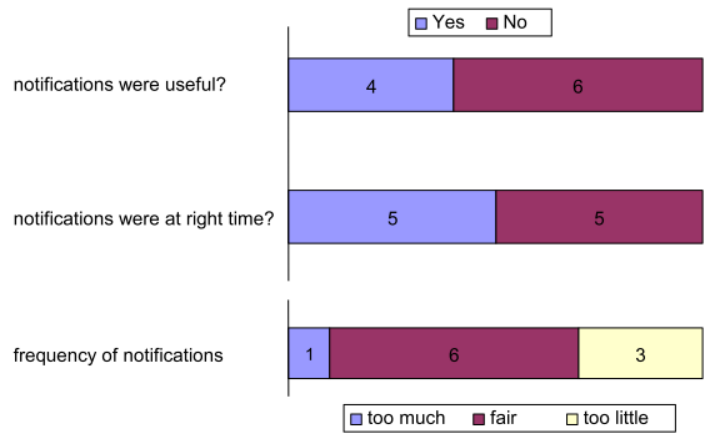

Figure 8. Content notification

However, more than 40% of the users mentioned the barrier notifications were useful as a whole. Therefore, although there is room for improvement, utility of this service that automatically notifies the proper content when the user is walking outside has been confirmed.

### 5.2.2. Content Submission

In this experiment, none of the users submitted a barrier or useful information. Major reasons would be difficulty of operation for the first-time users and the relatively short distance the user has walked in the experiment. Then, it resulted in a lack of interest to submit the information. So we need to consider a way to easily submit the content and to raise the motivation.

However, more than 70% of the users expressed a desire to submit the content in the questionnaire. Therefore, we confirmed that the elder people are willing to cooperate with each other.

### 5.2.3. User Interface

Some users commented that the maps displayed on the phone were not easy to see. And some users commented that the operations are generally not difficult, as the operation for the content retrieval involves just pushing a button (Fig. 9).

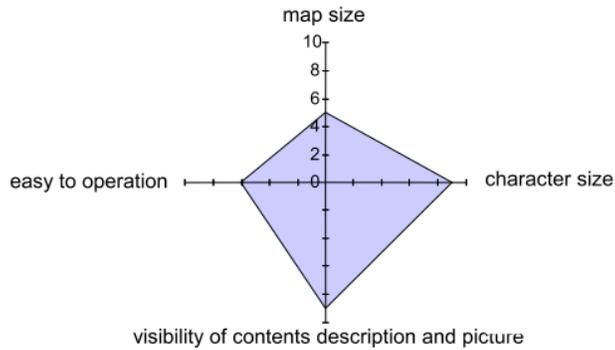

Figure 9. User Interface

However, the large images and large characters were appreciated, and the content was readily understood by the elderly.

### 5.2.4. Users' Impression

Table 2 shows the content which the users considered useful. In total, 90% of the users mentioned that this service is fun and they would like to use it again. However, there was a precondition for the continuous use, which is easier operation overall.

Table 2. Users' requests

| Content useful for notification | request for notification |
|---|---|
| police box | street people and crows |
| benches in the shade | automatic vending machine |
| park map | toilets |
| | resting places |
| | restaurants |

We observed that the user's route selection is different depending on the user's familiarity with that area. For example, the user who is familiar with the area, walked on side roads in the shade, but the unfamiliar users walked on the main street under the sun. Furthermore, we confirmed the barriers can be classified into two categories: static barriers and dynamic ones (Table 3). The former are structural barriers such as stairs and steps. On the other hand, the latter are changing dynamically with the passage of time. For example, the users who walked on the shady side of the street changed their route in the evening, because street people and crows appeared as it became darker. These observations highlighted the importance of selecting notification content according to the user profile and time. In the next section, we evaluate the accuracy of content notification.

Table 3. Static and dynamic barriers

| Static barrier | Dynamic barrier |
|---|---|
| steep stairs | crowd in station |
| pedestrian bridge | bicycles on sidewalk |
| road without sidewalk | road construction |
| no resting place | children in public space |
| steps, stairs | road under the sun |
|  | space without people in the night |
|  | street people, crows |
|  | hawkers |

## 6. EXPERIMENT ON ACCURACY OF CONTENT SELECTION

### 6.1. Experiment Setting

We conducted the second experiment to evaluate our content selection mechanism which corresponds to the above requirement 1. Firstly, in cooperation with the elder users, we collected 1200 data composed of the user profiles, the environmental parameters, the barrier information, and 2 or 3 possible reactions as a data set (Table 4). Then, we had 3-fold cross validation using the Bayesian network system, BayoNet [13]. In this experiment, any barrier for which the reaction determined "neglect" was considered to be the unnecessary content, and not notified. The other content was notified together with possible reactions.

Table 4. Example of data set

| Weather | T | Locality | Willingness to walk | Barrier | Reaction |
|---|---|---|---|---|---|
| Fine | 30C+ | No | not walk | Bicycles on street | proceed with caution |
| Cloudy | 5C- | Yes | walk for exercise | stairs in station | escalator |
| Rain | other | No | walk for exercise | Bicycles on sidewalk | detour |
| Fine | other | Little | walk for exercise | stairs in station | neglect |
| Fine | 5C- | Little | not walk | crowd in station | change time slot |
| Cloudy | other | Little | not walk | road w/o sidewalk | proceed with caution |
| Cloudy | 5C- | Yes | walk for exercise | street people | detour |
| Rain | 5C- | Little | not walk | stairs in station | elevator |

### 6.2. Experiment Result

The result is shown in Table 5. The average accuracy was 78.5%, and it can be said that the certain degree of accuracy is archived, because the reactions for each barrier has 2 or 3 choices (including "neglect"), so that the expectation was 41%.

In detail, the reaction for the illegally-parked bicycles remained 50% accuracy, because the reactions of the user vary and could not be modelled well in the data set. The same thing happened in case of homeless people and crows. On the other hand, the reaction for the static barrier such as stairs marked higher accuracy.

As lessons learnt, in addition to building a larger-scale of data set, we should consider personalization of the content selection in the future. If we compare the proposed reaction and the actual action of the user which can be obtained from service log like GPS, we will be able to reflect the user's preference such as "go up stairs for health" or "get on escalator" to the future notification. Moreover, the notifications at the frequently-visiting area may be limited to the dynamic or temporary barriers.

Table 5. Result of content selection

|  | T | F | Accuracy |
|---|---|---|---|
| 1st | 315 | 85 | 78.8% |
| 2nd | 316 | 84 | 79.0% |
| 3rd | 311 | 89 | 77.7% |
| Ave. |  |  | 78.5% |

## 7. CONCLUSION

In this paper, we proposed a mobile service for a GPS cellular phone to support the elderly go out into society. Then, we described two experiments on the usability and the accuracy of the content notification by the Bayesian network. Then, we confirmed that this service is useful for the elderly people in terms of its notification feature, although the interface needs some improvements.

We are currently building a larger set of the data to improve the content notification mechanism and a service logging mechanism for the personalization, so that we will be able to promote the participation of the elderly in social activities. Also, this service would be useful for handicapped people and women having a baby with minor modification.

**Authors**

**Keisuke Umezu** is a researcher at the Knowledge Discovery Research Laboratories, NEC Corp. He received his Master degree from the University of Electro-Communications, Japan.

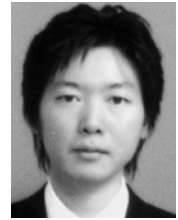

**Takahiro Kawamura** is a Senior Research Scientist at the Corporate Research and Development Center, Toshiba Corp., and also an Associate Professor at the Graduate School of Information Systems, the University of Electro-Communications, Japan.

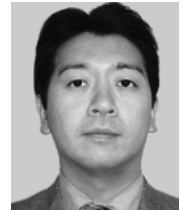

**Akihiko Ohsuga** is a Professor at the Graduate School of Information Systems, the University of Electro-Communications. He is currently the Chair of the IEEE Computer Society Japan Chapter.

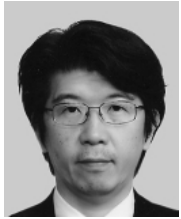